# Experimental Study on Time Series Analysis of Lower Limb Rehabilitation Exercise Data Driven by Novel Model Architecture and Large Models

Hengyu Lin, Shideng Ma, Zixun Luo

**ABSTRACT**

This study investigates the application of novel model architectures and large-scale foundational models in temporal series analysis of lower limb rehabilitation motion data, aiming to leverage advancements in machine learning and artificial intelligence to empower active rehabilitation guidance strategies for post-stroke patients in limb motor function recovery. Utilizing the SIAT-LLMD dataset of lower limb movement data proposed by the Shenzhen Institute of Advanced Technology, Chinese Academy of Sciences, we systematically elucidate the implementation and analytical outcomes of the innovative xLSTM architecture and the foundational model Lag-Llama in short-term temporal prediction tasks involving joint kinematics and dynamics parameters. The research provides novel insights for AI-enabled medical rehabilitation applications, demonstrating the potential of cutting-edge model architectures and large-scale models in rehabilitation medicine temporal prediction. These findings establish theoretical foundations for future applications of personalized rehabilitation regimens, offering significant implications for the development of customized therapeutic interventions in clinical practice.

## 1. INTRODUCTION

With the intensification of population aging and rising prevalence of chronic diseases, global healthcare systems and rehabilitation frameworks are confronting unprecedented challenges. Stroke – ranked as the second leading cause of death (accounting for 11.6%[10.8–12.2] of total mortality) and the third largest contributor to disability-adjusted life years (5.7%[5.1–6.2])[3] – is an acute cerebrovascular event characterized by sudden brain vessel rupture or obstruction, resulting in cerebral ischemia or hemorrhage. Post-stroke conditions frequently trigger chronic health complications, including hemiplegia, dysphagia, cognitive impairment, and depression. From 1990 to 2019, stroke incidence surged by 70.0%(67.0–70.0), prevalence escalated by 85.0%(83.0–88.0), and mortality increased by 43.0%(31.0–55.0) [3], underscoring the urgency of integrating post-operative rehabilitation with chronic disease management.

Active patient engagement in limb motor rehabilitation is critical for leveraging movement intention to induce neuroplasticity and enhance recovery efficiency, which serves as the cornerstone for restoring activities of daily living(ADL) and reducing long-term care dependency[1][2]. Motor intention signals are typically categorized into biological and non-biological modalities[20]. Among these, surface electromyography(sEMG) signals have emerged as a promising approach for motion intention prediction due to their non-invasiveness, acquisition stability, and signal fidelity. Conventional methodologies employ filtered sEMG root mean square(RMS) features as inputs to predict joint kinematics(angular trajectories) and dynamics(torque profiles) [21][22], where continuous lower-limb gait cycles can be formulated as short-term multivariate temporal forecasting tasks.

The integration of artificial intelligence (AI) with smart healthcare represents a pivotal frontier in technological advancement. While neural architectures – including recurrent neural networks[5][22], convolutional neural networks[6], hybrid LSTM architectures[7][23], attention-based encoder-decoders [8], graph neural networks[9], and Transformers[10] – have demonstrated competitiveness in temporal forecasting, significant challenges persist in handling nonlinear, multimodal, and high-dimensional data. These limitations stem from: (1)short-sequence temporal dependencies in most clinical datasets; (2)complex interdependencies between multi-scale temporal features; and (3)insufficient model interpretability[11][12][13][14][15].

Notably, the rapid evolution of foundation models (FMs), empowered by breakthroughs in large language models (LLMs), has introduced a paradigm shift in representation learning[16]. By synergizing the flexibility of deep learning architectures with FM-driven approaches, this study experimentally investigates: the feasibility of novel model architectures (e.g., xLSTM) in constructing foundational frameworks for kinematic/dynamic temporal prediction; and the applicability of large-scale pre-training and fine-tuning paradigms in lower-limb rehabilitation scenarios.

## 2. RELATED WORK

This section delineates the analytical framework for existing public datasets in lower-limb rehabilitation medicine and motion intention recognition contexts, accompanied by formal documentation of the selected dataset (see Section 2.1). Furthermore, it elucidates the rationale for adopting xLSTM[40](Section 2.2) and Lag-Llama[46] (Section 2.3) as representative novel architectures and foundation model-driven paradigms, respectively, highlighting their technical superiority in lower-limb gait temporal forecasting tasks.

### 2.1. Dataset: SIAT-LLMD

The advent of non-invasive surface electromyography (sEMG) acquisition technologies has catalyzed rapid advancements in motion intention and state prediction for rehabilitation applications [24][25]. For upper limb studies, the Ninapro database [26][27] – a multimodal public dataset comprising over 180 recordings from intact-limb and transradial amputee subjects – provides comprehensive electromyographic, kinematic, inertial, clinical, neurocognitive, and eye-hand coordination data, establishing itself as a benchmark resource for machine learning research in human-robot prosthetics. In contrast, existing lower-limb motion intention datasets exhibit notable limitations: E.Reznuck [28] and RBD [29] lack sEMG measurements; HAR-SEMG [30] omits synchronized kinematic (joint angles) and dynamic (joint torque) parameters; while J.Camargo and HuMoD datasets [31][32], though addressing prior gaps, require computationally intensive gait phase feature extraction.

The Shenzhen Institute of Advanced Technology Lower Limb Motion Dataset (SIAT-LLMD), developed by Wenhao Wei et al. [33], overcomes these deficiencies through multimodal integration. Utilizing motion capture systems, six-axis force platforms, and wireless sensors positioned on thigh/calf muscles, it captures raw 9-channel sEMG signals alongside synchronized kinematics and dynamics data from 40 healthy subjects performing 16 motor tasks.

**2.2. xLSTM**

The evolution of neural networks in temporal sequence analysis has progressed through distinct developmental phases. Initially, standard feedforward neural networks demonstrated limited efficacy in capturing long-range dependencies within temporal data due to their inherent lack of memory retention mechanisms. This limitation prompted the development of recurrent neural networks (RNNs) [34], which introduced cyclic computational structures to preserve historical information. However, RNNs suffered from the vanishing gradient problem[35], where error gradients diminish exponentially during backpropagation through deep layers, fundamentally constraining their capacity to model extended temporal relationships.

To address this critical challenge, Hochreiter and Schmidhuber[36] pioneered the Long Short-Term Memory (LSTM) architecture in 1997, incorporating sophisticated gating mechanisms (input/forget/output gates) to regulate information flow. This innovation effectively mitigated gradient dissipation while enabling adaptive memory management[37], establishing LSTM as a preferred architecture for lower-limb motion intention recognition tasks. Nevertheless, opportunities for architectural optimization persist in enhancing long-term dependency modeling and computational adaptability [38][39][42].

Recent advancements by Maximilian Beck et al. introduced xLSTM[40], an enhanced architecture for multivariate long-term time series forecasting (LTSF) that integrates exponential gating mechanisms with augmented memory capacity through modified cell states. Complementary research by Musleh Alharthi and Ausif Mahmood demonstrated xLSTMTime's [41] superior predictive performance against state-of-the-art benchmarks across diverse real-world datasets. These technical merits motivate our experimental selection of xLSTM as a representative novel architecture to systematically evaluate its applicability in temporal analysis tasks for lower-limb motion intention recognition and rehabilitation therapeutics.

**2.3. Lag-Llama**

The expansion of neural architectures, coupled with the exponential growth in parameter scale and training corpus of large language models (LLMs), has enabled medical-domain LLMs to demonstrate promising capabilities through few-shot fine-tuning[43]. Nevertheless, current research has yet to extensively validate these capabilities in physiological and behavioral temporal analysis tasks requiring complex sequence modeling.

Subsequent research by Xin Liu et al.[44] substantiated that strategic prompt engineering and parameter-efficient fine-tuning (PEFT) on 24-billion-parameter LLMs can effectively leverage temporal data for cardiovascular, metabolic, and physiological health predictions. Furthermore, the open-source foundation model Lag-Llama[46] – built upon the decoder-only Transformer architecture LLaMA [45] while incorporating lagged features as covariates – has exhibited state-of-the-art zero-shot performance and few-shot adaptation capabilities in univariate probabilistic time series forecasting. In subsequent experimental chapters, we employ Lag-Llama as the representative foundation model-driven framework for empirical validation in lower-limb motion intention recognition and rehabilitation therapeutics.

## 3. EXPERIMENT

In this section, we will perform preprocessing, feature extraction, and feature normalization operations on target data (DNS, UPS) from the SIAT-LLMD dataset (See Section 3.1). Gaussian Process Regression analysis will be conducted to verify the completeness of cleaned data (See Section 3.2). Subsequently, using the dataset as input, we will test the applicability of the xLSTMTime architecture in predicting joint-related angles and torque from sEMG signals (See Section 3.3). Finally, we will evaluate and compare the forecasting performance of Lag-Llama on time series data under both zero-shot and parameter fine-tuning configurations (See Section 3.3).

### 3.1. Data Processing Pipeline

The raw sEMG signals labeled DNS and UPS from the SIAT-LLMD dataset (Figure 1) undergo systematic preprocessing and feature extraction through the following computational workflow:

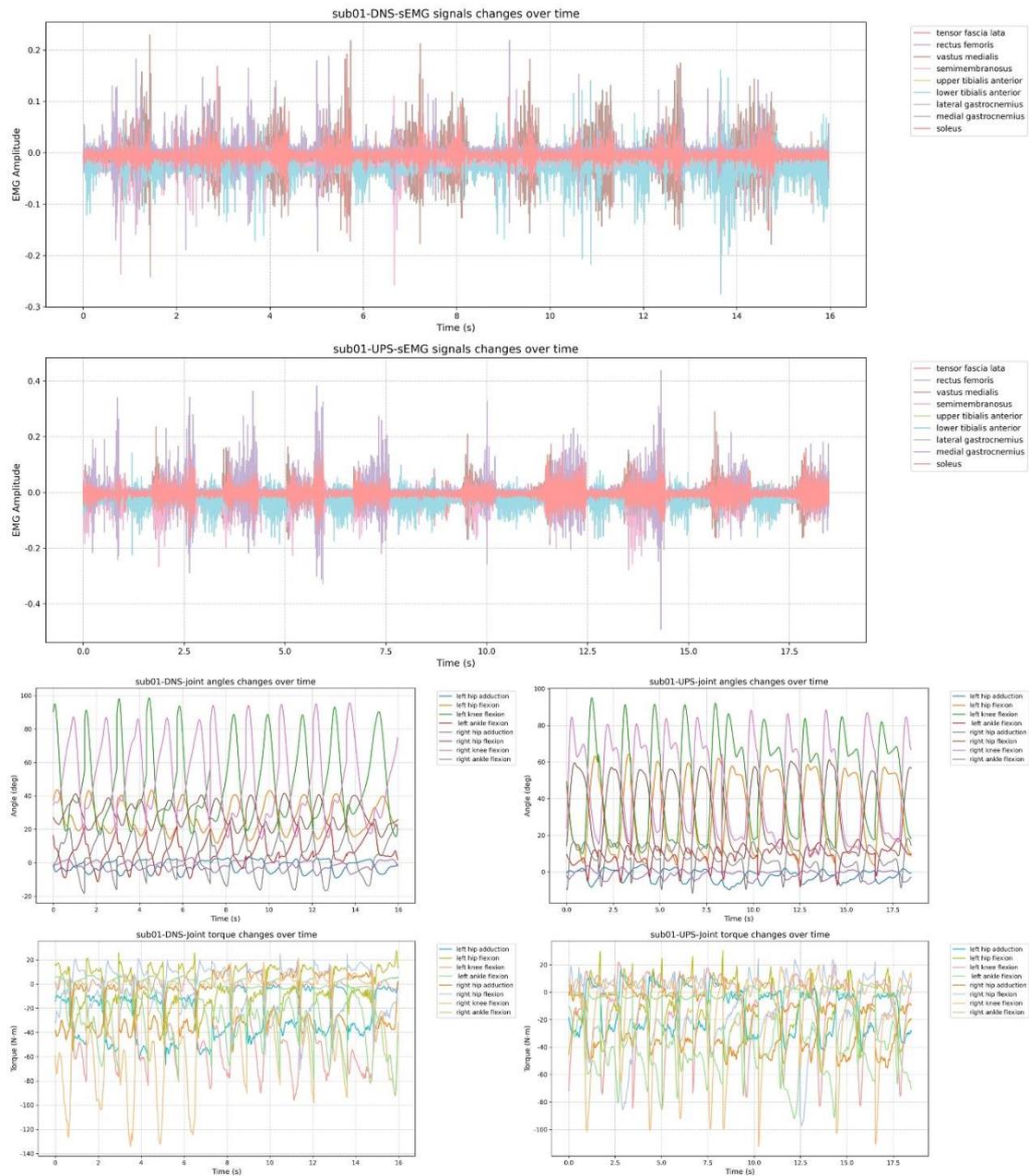

Figure 1. Temporal Variation Distribution of sEMG Signals, Joint Angles, and Joint Torque Data During DNS and UPS Gait Conditions in Subject 01

First, the EMG signals undergo baseline correction, wavelet packet transform, and Butterworth filtering. The wavelet packet transform employs a threshold of 0.08 with a decomposition level of 8, while the Butterworth filter is initialized with a 7th-order configuration. A processing window size of 100 data points is defined with a 50-point overlap between consecutive windows. This overlapping window design mitigates boundary effects, such as spectral leakage in Fourier transform applications. Subsequently, L1-norm-based maximum absolute value normalization is applied to the processed data. For feature extraction, six parameters are calculated from the overlapping windows: integral, variance, wavelength, zero-crossing rate, correlation coefficient, and weighted average frequency. All extracted features undergo standardized processing (Figures 2 and 3).

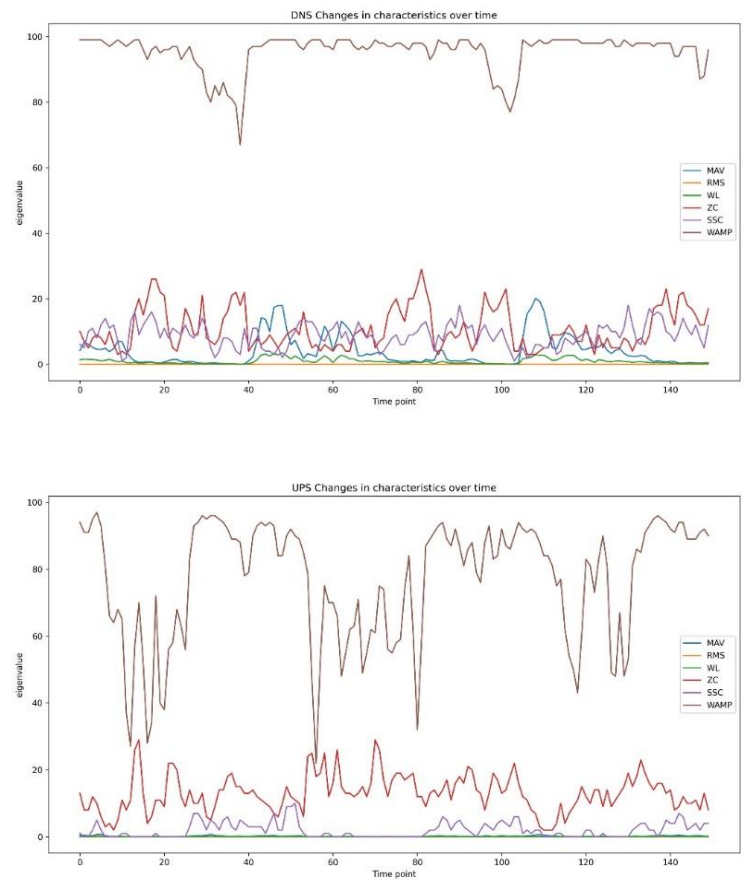

Figure 2. Temporal Distribution of sEMG Features Across the First 150 Time Points During DNS and UPS Conditions

Following the aforementioned processing, the DNS gait yielded an sEMG feature matrix of dimensions 22551×9×6 and a corresponding joint kinematics dataset of 22551×8×2, while the UPS gait produced an sEMG feature matrix of 19952×9×6 and a joint dataset of 19952×8×2. The sEMG matrices (x×9×6) represent sample quantity (x), 9 electrode channels, and 6 extracted features (integral, variance, wavelength, zero-crossing rate, correlation coefficient, weighted average frequency), respectively. The joint datasets (x×8×2) encompass bilateral hip adduction, hip flexion, knee flexion, and ankle flexion (4 joints per leg × 2 legs), with the final dimension quantifying joint angles (°) and torque (Nm). The types of feature normalization and extraction are detailed in Figure 2.

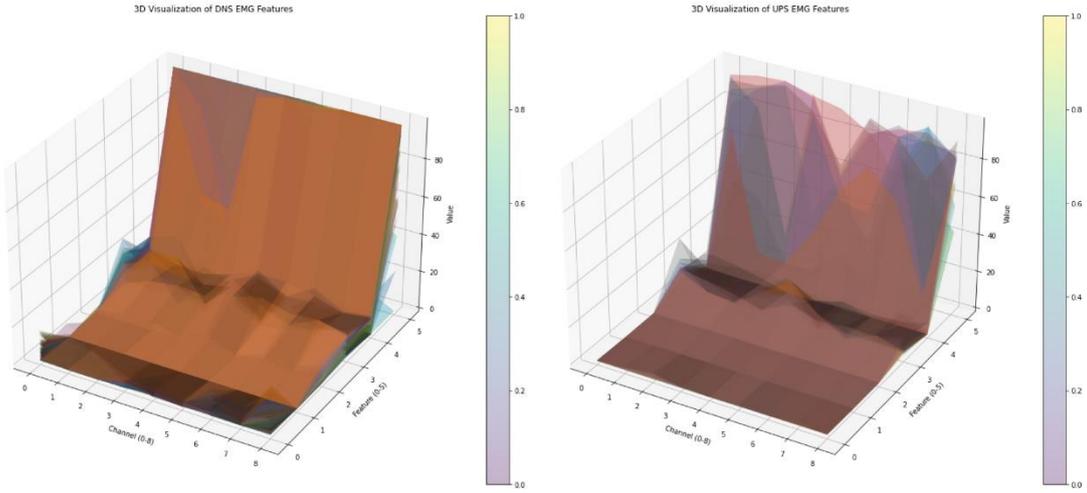

Figure 3. Three-Dimensional Visualization of Extracted sEMG Features Under DNS (Left Panel) and UPS (Right Panel) Gait Conditions

### 3.2. Gaussian Process Regression

Gaussian Process Regression (GPR) is a nonparametric Bayesian regression methodology that computes the posterior distribution of target functions given input observations, where the mean function maps predictive estimates and the covariance quantifies prediction uncertainty. This study employs GPR to validate the efficacy of sEMG-derived features in predicting lower-limb joint kinematics (angular trajectories) and dynamics (torque profiles) during gait cycles, thereby establishing a theoretical foundation for sEMG-driven biomechanical modeling.

The formal definition of a GPR kernel function is:

$$k(\mathbf{x}, \mathbf{x}') = \sigma_f^2 \cdot \exp\left(-\frac{\|\mathbf{x} - \mathbf{x}'\|^2}{2\ell^2}\right)$$

with hyperparameter bounds:

$$\sigma^2{}_f \in [10^{-3}, 10^3], \quad l \in [10^{-2}, 10^2]$$

Gaussian Process Regression (GPR) models were independently developed for joint angle and torque prediction, utilizing the extracted feature matrices and kinematic/torque datasets as inputs. The DNS gait achieved a mean absolute error (MAE) of 6.28463 and root mean squared error (RMSE) of 10.54293 for joint angle prediction, while joint torque prediction yielded an MAE of 7.68814 and RMSE of 11.81379. For UPS gait, joint angle prediction exhibited an MAE of 7.54123 and RMSE of 12.28001, with torque prediction showing higher deviations (MAE: 10.95209, RMSE: 16.21753). The predictive performance of the DNS-derived model on UPS gait kinematics is further visualized in Figure 4.

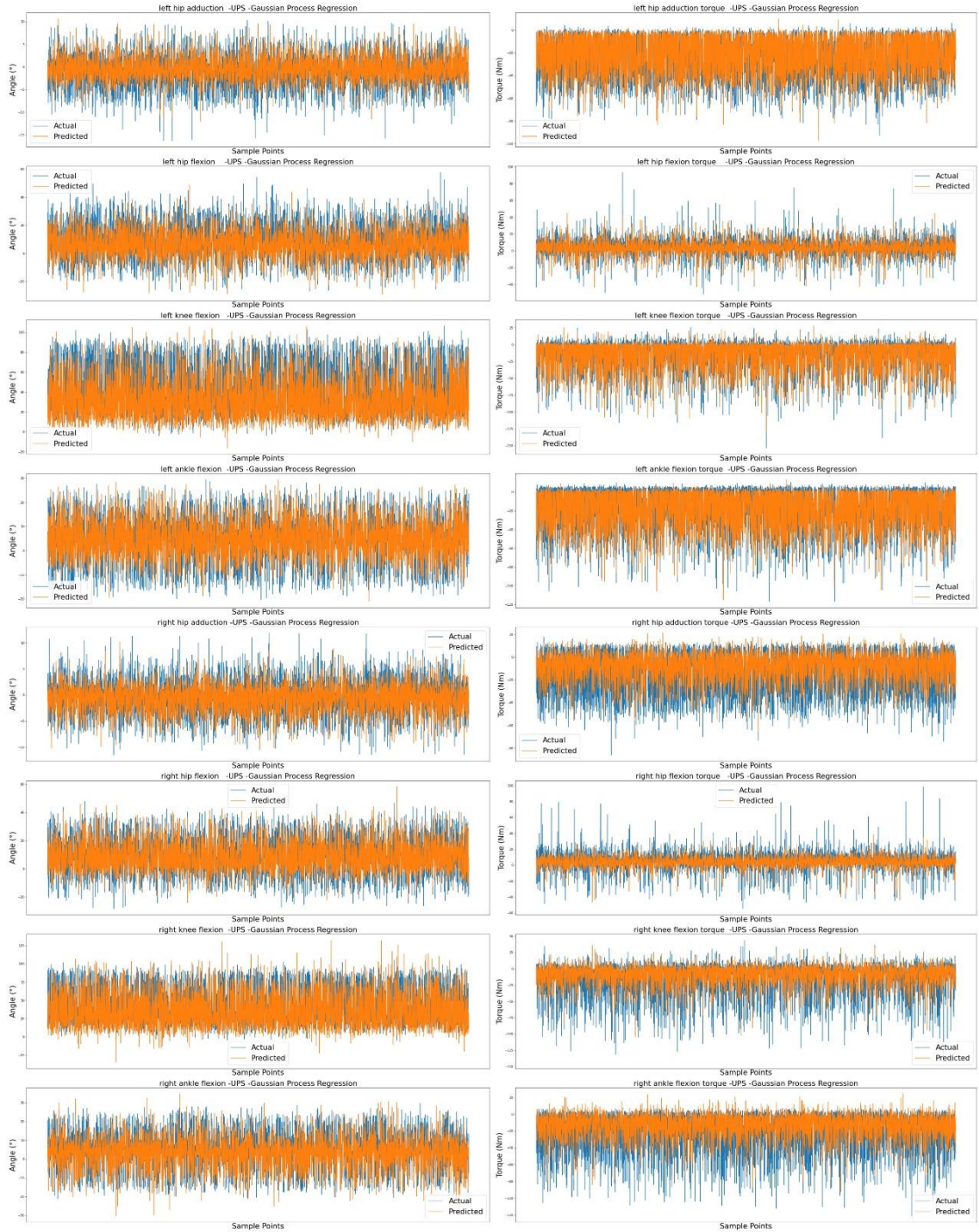

Figure 4. Gaussian Process Regression Modeling and Prediction of Lower-Limb Joint Angles and Torque During UPS Gait

## 3.3. xLSTM Model Training

xLSTM addresses the limitations of traditional LSTMs in handling large-scale data and long sequences by introducing novel gating mechanisms and memory architectures to improve poor parallelism and insufficient memory capacity. The exponential gating mechanism provides more dynamic information filtering capabilities, enhancing controllability and flexibility in information flow regulation. Simultaneously, the integrated novel memory mixing structure (sLSTM) and matrix memory (mLSTM) effectively enhance both storage capacity and utilization efficiency. Furthermore, the incorporation of normalization, stabilization techniques, and residual blocks ensures xLSTM maintains operational stability while efficiently processing complex sequential data. These architectural advancements demonstrate xLSTM's potential advantages in predicting multidimensional time-series data for lower limb gait analysis.

The fundamental components primarily include the Causal Convolutional Layer (CausalConv1D), which ensures that the current output depends only on past inputs, and the BlockDiagonal transformation, which partitions the input into multiple heads, each of which independently performs a linear transformation.

For sLSTM (structured LSTM), the primary implementation involves stabilizing training through normalization techniques such as LayerNorm and GroupNorm, incorporating gating mechanisms and causal convolutions to process temporal information, and employing projection factors for dimensionality transformation. For mLSTM (multiplicative LSTM), the core design utilizes attention mechanisms (query, key, value), integrates skip connections, and leverages multiplicative interactions to enhance feature representation.

Given the electrode sampling frequency of SIAT-LLMD at 1926 Hz, the data samples were processed according to the sampling frequency. The input feature dimension and channel dimension were defined to align with the sample matrix shape. The model was initialized with a hidden layer size of 32, 2 layers, 20 training steps, and a learning rate of 0.01, with the corresponding RMSE loss curve illustrated in Figure 5.

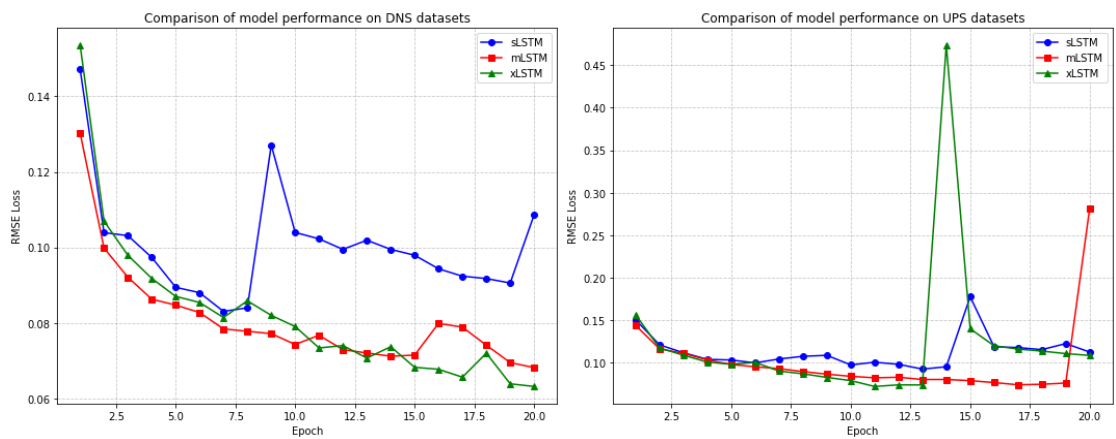

Figure 5 demonstrates the predictive performance of xLSTM and its sub-models on DNS and UPS.

From the experimental results, it can be observed that xLSTM, which combines the structural characteristics of sLSTM and the multiplicative interaction strengths of mLSTM, is well-suited for handling complex temporal patterns and nonlinear relationships, demonstrating superior modeling capabilities for the mapping between electromyographic (EMG) signals and joint movements. mLSTM exhibits stable performance on the DNS dataset but shows instability on the UPS dataset (with higher final RMSE). Its multiplicative interaction mechanism effectively captures nonlinear features in signals and achieves relatively faster convergence, as shown in Figure 6. In contrast, sLSTM delivers stable yet lower-precision performance, with a simpler architecture and reduced computational overhead, making it suitable as a baseline model.

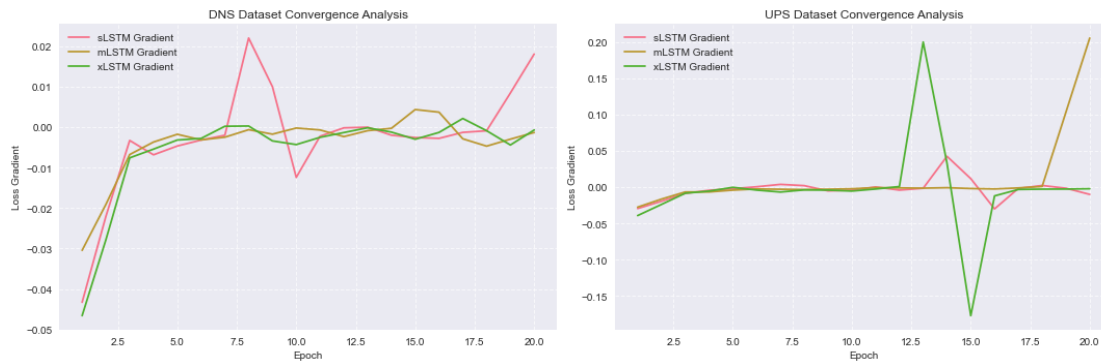

Figure 6 presents the convergence analysis of xLSTM and its sub-models on the DNS and UPS datasets.

The hybrid architecture integrating structured feature representation with multiplicative interaction mechanisms demonstrates compelling efficacy in decoding the complex relationships between bioelectric signals and human biomechanics, particularly within clinical rehabilitation scenarios requiring robust temporal pattern recognition and nonlinear mapping capabilities.

### 3.4. Lag-Llama Zero-Shot Prediction

Lag-Llama, designed for univariate probabilistic forecasting, employs a frequency-agnostic generic approach to tokenize time-series data, enabling its open design to generalize effectively to unseen frequencies. It leverages the Transformer architecture to parse input tokens and map them to future predictions with confidence intervals. The autoregressive process of generating predictions efficiently allows the model to generate prediction uncertainty intervals.

In terms of data processing, Normalize multidimensional biosignals (sEMG, joint angles, torque) to zero-mean/unit-variance.Reshape and concatenate samples into GluonTS-compatible format.Generate iso-sampled time indices ($\Delta t = 0.519$ ms) aligned with 1,926 Hz acquisition rate, and decouple target variables (angle/torque) into univariate series with shared temporal metadata

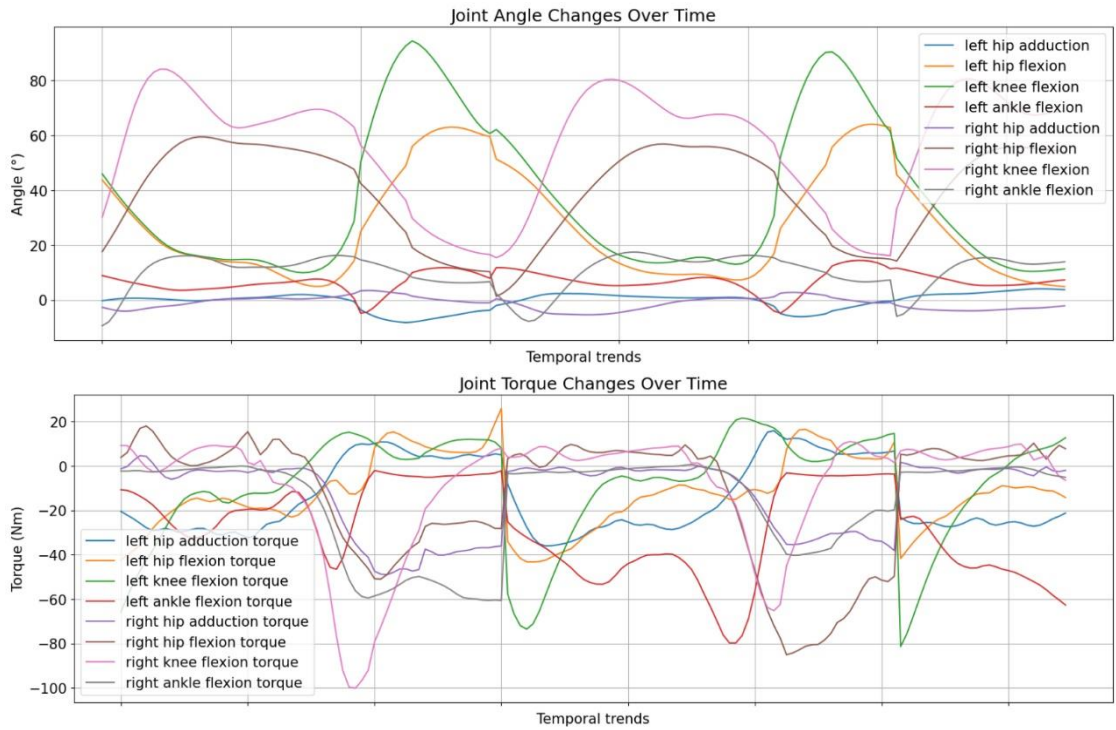

Figure 7 Reshaped DNS joint motion angles and joint torque time-series data

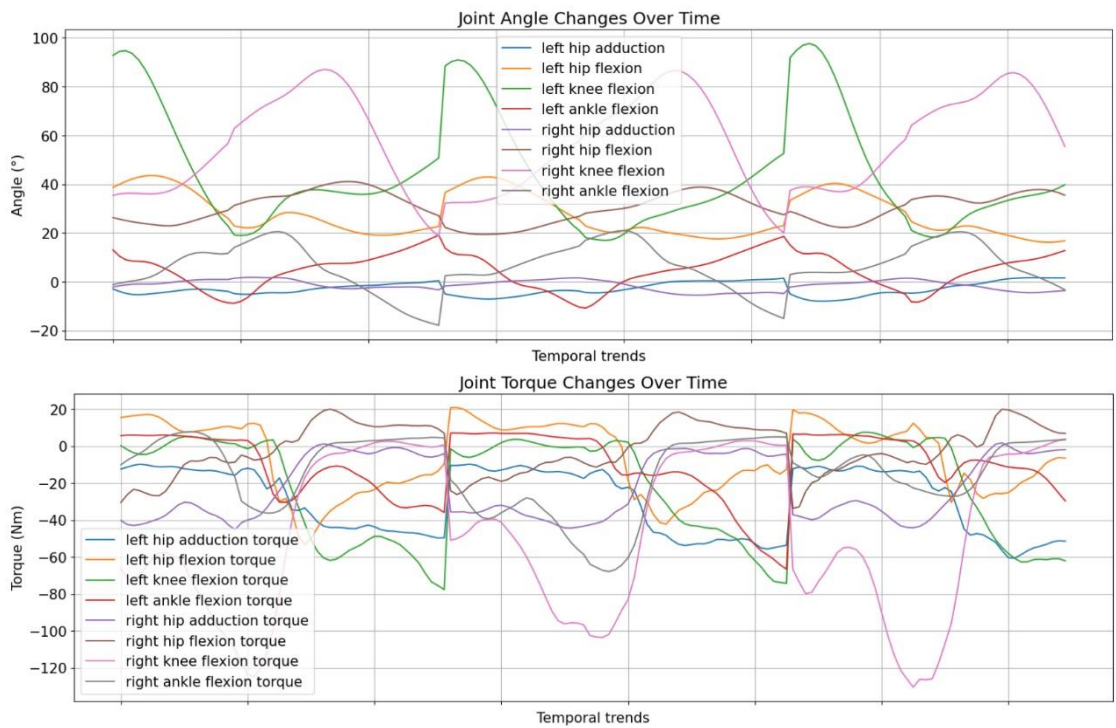

Figure 8 Reshaped UPS joint motion angles and joint torque time-series data

Configures a forecast horizon of 128 steps with a context window twice the horizon length (256 steps) to optimize temporal dependency modeling. In the zero sample prediction of joint Angle and torque, the DNS gait prediction results are shown in Figure 9, and the UPS gait prediction results are shown in Figure 10.

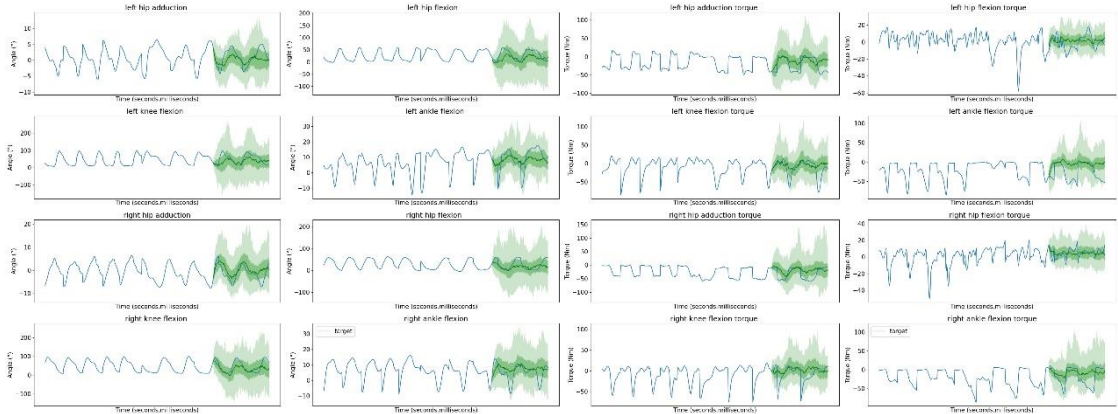

Figure 9 Zero-shot prediction results of Lag-Llama on the DNS dataset

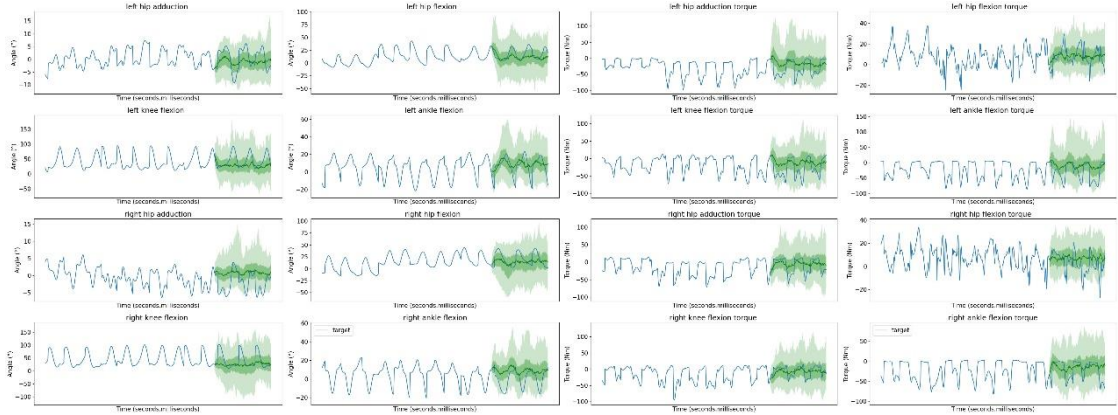

Figure 10 Zero-shot prediction results of Lag-Llama on the UPS dataset

From the results, it can be observed that the model partially captures the overall trend of joint torque variations, with the direction and magnitude of predicted fluctuations broadly aligning with the actual data. The uncertainty estimates in its output probability distributions (green shaded areas) hold significant application value and potential for assessing postoperative lower limb motor function recovery in patients. However, deviations between predictions and ground truth occur in rapidly changing regions, with notable discrepancies in fluctuation magnitudes for specific joints (e.g., the right ankle joint). Additionally, as the prediction horizon extends, the prediction uncertainty (width of the green shaded areas) increases significantly, indicating the model's limited reliability in long-term predictions.

Overall, the Lag-Llama model demonstrates its potential for zero-shot prediction of human joint torque in postoperative lower limb rehabilitation scenarios, particularly for short-term assistive medical rehabilitation applications.

### 3.5. Lag-Llama Fine-Tuning

Although the Lag-Llama base model has demonstrated promising potential in zero-shot prediction of human joint torque—particularly for short-term, multi-sample-point time-series tasks—further experimental validation of its adaptability is required to enhance prediction accuracy and align it with practical rehabilitation medicine requirements. To address this, we will conduct fine-tuning of the model tailored to the specific characteristics of the dataset. Through this process, we aim to elucidate the model's potential for performance improvement on task-specific sequential data and critically evaluate its effectiveness and limitations in real-world applications.

The study conducted benchmark testing on the performance of the fine-tuned model, attempting to extract optimal performance from the model. For the finalized model configuration, the key parameters included a prediction length of 128, a context length of 512, with model architecture parameters loaded from a checkpoint. Fine-tuning training employed an early stopping strategy over 50 epochs.

The prediction results for joint angles and torque after fine-tuning are illustrated in Figures 11 and 12 (Figure 11: DNS; Figure 12: UPS).

From the preliminary observation of the prediction confidence intervals, it is evident that most probability distributions have significantly converged after initial fine-tuning, aligning closely with the data target line (blue) and exhibiting narrow spreads. This indicates a positive outcome, suggesting that the fine-tuned model has largely captured the underlying trends and patterns of the data. However, deviations emerge in predictions for later time points, such as in left hip joint flexion angles and right hip joint flexion torque.

In probabilistic time-series forecasting, the Continuous Ranked Probability Score (CRPS) serves as a metric to evaluate the accuracy of probabilistic predictions by quantifying the discrepancy between the predicted distribution and the observed values. A lower CRPS value indicates superior predictive model performance.

In our results, the mean CRPS for joint motion angle predictions was 0.5161 with a standard deviation of 0.1140, while for joint torque predictions, the mean CRPS was 0.8430 with a standard deviation of 0.2571. The box plots illustrating the CRPS distributions for the fine-tuned Lag-Llama predictions of joint motion angles and torque are presented in Figure 13.

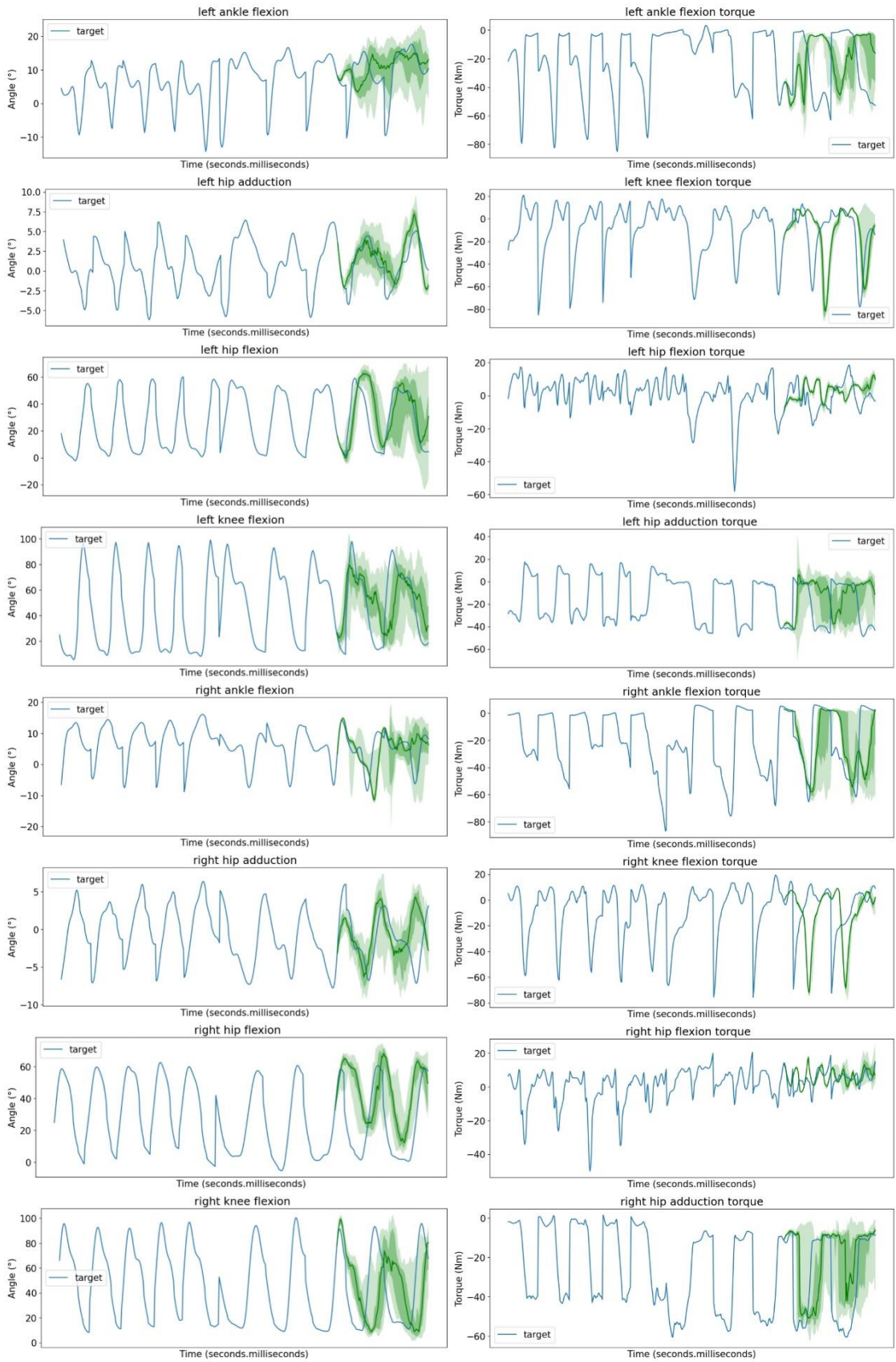

Figure 11 Prediction results of the fine-tuned Lag-Llama model on DNS gait

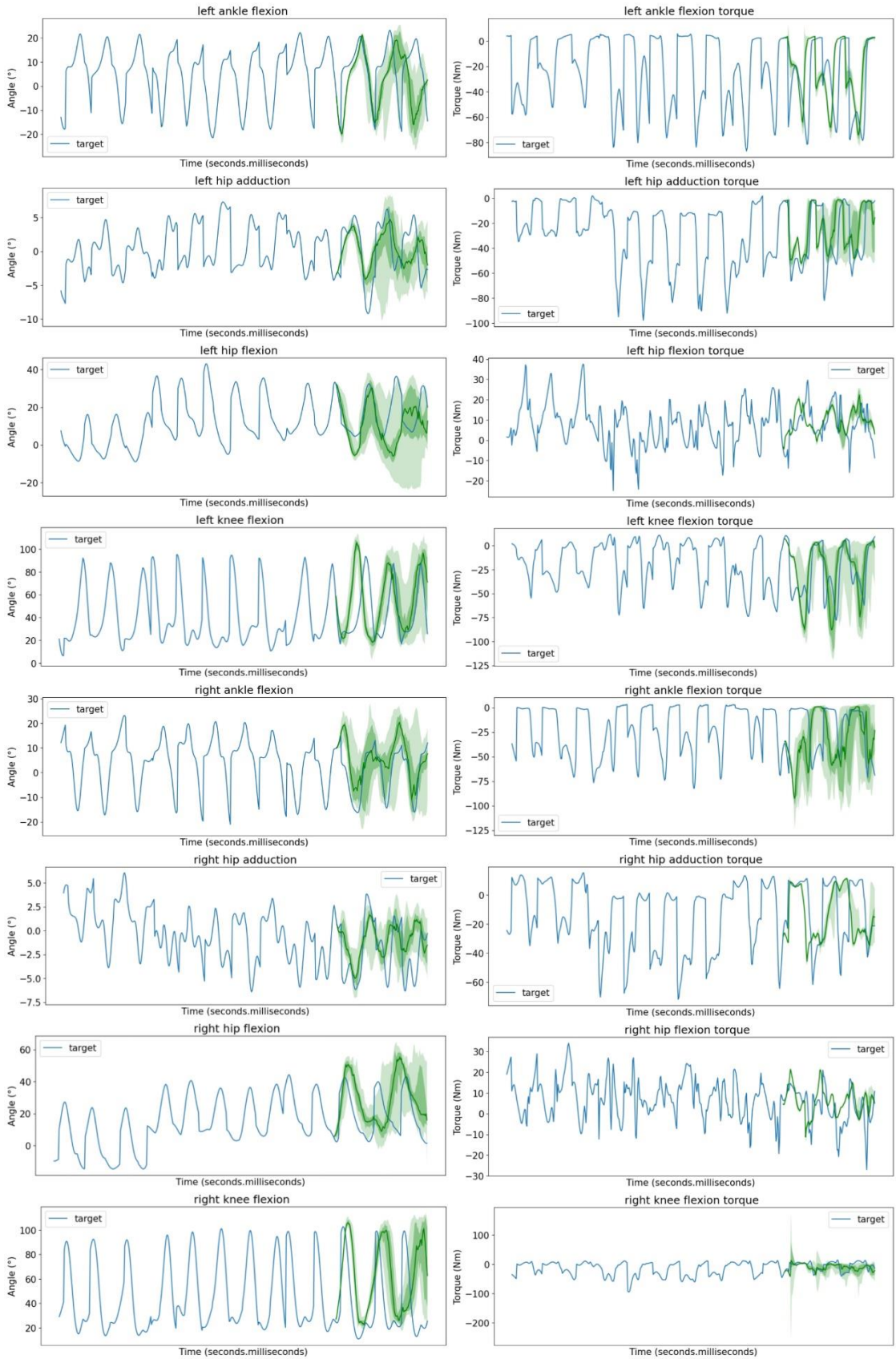

图 Figure 12 Prediction results of the fine-tuned Lag-Llama model on UPS gait

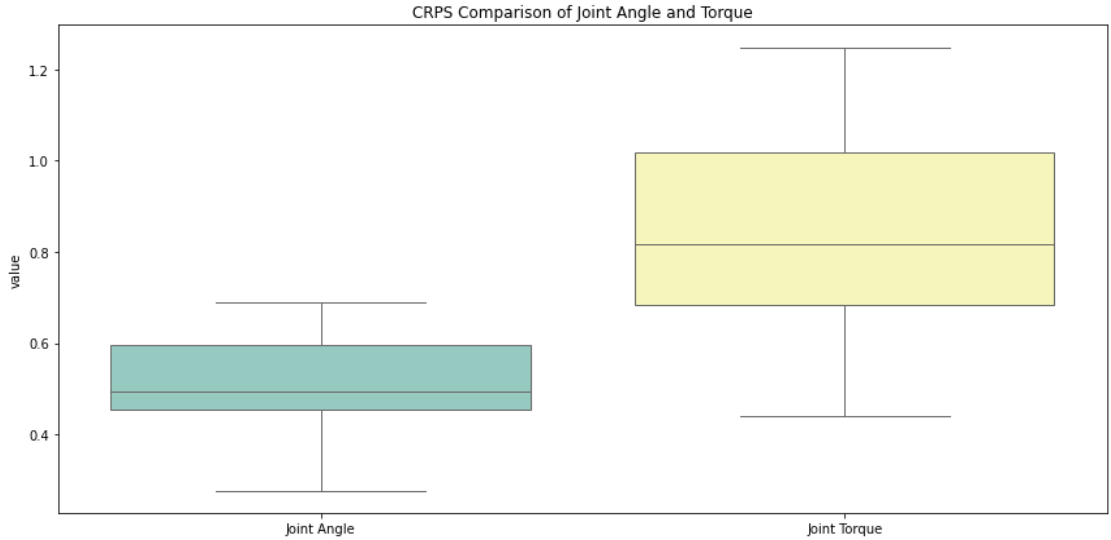

Figure 13 Comparison of CRPS (Continuous Ranked Probability Score) for joint angles and torque

Overall, the CRPS values for joint torque are consistently higher than those for joint angles, indicating greater prediction errors for torque under the current model. However, the broader distribution of joint torque data also demonstrates a degree of generalizability. In the context of short-term time-series forecasting for lower limb gait motion studied here, the model exhibits relatively superior performance in predicting joint angles, while the prediction accuracy for joint torque shows larger errors and significant optimization potential, likely attributable to the inherent complexity and variability in torque dynamics.

## 4. RESULTS

This study explores the application of novel model architectures and large-scale models in time-series analysis of lower-limb rehabilitation motion data, providing new insights to enhance rehabilitation efficiency. The results demonstrate that xLSTM and Lag-Llama exhibit promising potential in predicting joint motion angles and torque from short-term time-series data samples, offering valuable references for rehabilitation therapy.

The xLSTM model demonstrates notable accuracy in predicting both joint angles and torque. By integrating the structural strengths of sLSTM and the multiplicative interaction advantages of mLSTM, it performs effectively in modeling the complex relationship between bioelectrical signals and human motion. The mLSTM model achieves stable performance on the DNS dataset but exhibits instability on the UPS dataset, with its multiplicative interaction mechanism facilitating the capture of nonlinear signal features and relatively faster convergence. The sLSTM model, while less precise, delivers stable performance with simpler architecture and lower computational overhead, making it suitable as a baseline model.

The Lag-Llama model shows potential in zero-shot prediction, capturing the overall trends of joint torque variations, with predicted fluctuation directions and magnitudes broadly aligning with actual data. The uncertainty estimates in its output probability distributions hold significant application value and prospects for assessing lower-limb motor function recovery in patients. After fine-tuning, the prediction performance of Lag-Llama improves, with most probability distributions converging significantly and demonstrating better accuracy in joint angle predictions. However, prediction errors for joint torque remain substantial.

Future research aims to further optimize existing model architectures and training methods to enhance task-specific performance, acquire targeted limb motion datasets (e.g., postoperative rehabilitation state samples) to support real-world clinical applications, and focus on large model-driven optimization strategies such as online learning and incremental learning to enable personalized rehabilitation plans for patients. Additionally, the exploration of model interpretability methods is critical, as it helps researchers and clinicians comprehend the model's predictions, thereby facilitating safer and more effective integration into rehabilitation therapies.

To ensure full reproducibility, methodological transparency, and open scientific collaboration, all experimental code, processed datasets (including anonymized DNS/UPS gait sequences), model configurations (encompassing xLSTM variants and Lag-Llama architectures), training pipelines, and comprehensive documentation associated with this research will be publicly accessible via GitHub repository (https://github.com/LINHYYY/TSA-LL-Rehab-xLSTM-LagLlama). These resources are provided under an open-source license to facilitate independent validation, extension of the proposed frameworks, and community-driven advancements in AI-assisted rehabilitation analytics.

## Acknowledgements

This work was supported by the following grants: Guangdong Provincial University Achievement Transformation Center under the Character-istic Innovation Research Project of Universities in Guangdong Province (Grant No. 2022DZXX02), titled "Collaborative Motion Control of Agents (Robots) in Edge Computing Mode".

## Author contributions

Hengyu Lin conceived and designed the research, conducted experiments, processed and analyzed data, interpreted results, and wrote the manuscript; Shideng Ma secured research funding, oversaw the overall project design and coordination; Zixun Luo optimized data visualization and verified the accuracy of the data and results.

## Competing interests

The authors declare no competing interests.